\documentclass[11pt]{article}

\usepackage{sectsty}
\usepackage{graphicx}
\usepackage{amsmath,amssymb}
\usepackage{bm}
\usepackage{hyperref}
\usepackage{multirow}
\usepackage{authblk}
\usepackage{url}
\usepackage{xcolor}
\usepackage{caption}
\usepackage{subcaption}

\usepackage[square,numbers]{natbib}

\newcommand{\rvec}{ \mathbf{r} }

\newcommand{\fvec}{ \mathbf{f} }

\newcommand{\nhat}{ \mathbf{\hat{n}} }
\newcommand{\barpsi}{ \bar{\psi} }

\title{Active particles confined in deformable droplets}
\author[1,2]{ Javier Diaz}
\author[1,2]{Ignacio Pagonabarraga}
\affil[1]{Departament de F\'isica de la Mat\`eria Condensada, Universitat de Barcelona, Mart\'i i Franqu\'es 1, 08028 Barcelona, Spain}
\affil[2]{Universitat de Barcelona Institute of Complex Systems (UBICS), Universitat de Barcelona, 08028 Barcelona, Spain}

\begin{document}
\maketitle

\begin{abstract}
 Active particles under soft confinement such as droplets or vesicles present intriguing phenomena, as collective motion emerges alongside the deformation of the environment. 
A model is employed to systematically investigate droplet morphology and particle distribution in relation to activity and concentration,revealing that active particles have the capacity to induce enhanced shape fluctuations in the droplet interface with respect to the thermal fluctuations, aligning with recent experimental observations. 
A rich phase behaviour can be identified with two different mechanism of droplet breakage.
\end{abstract}


\section{Introduction} 
Active matter, composed by building blocks that consume energy from the medium to perform work, represents a fascinating class of systems at the intersection of physics, chemistry, and biology. 
Active particles (APs), such as self-propelled microorganisms or synthetic nanoscale motors, exhibit intriguing behaviors in complex environments\cite{bechinger_active_2016}, including the emergence of collective motion under hard confinement\cite{woodhouse_spontaneous_2012,wioland_confinement_2013,lushi_fluid_2014}, in the presence of obstacles\cite{reichhardt_pattern_2023}, flow  guidance\cite{wioland_ferromagnetic_2016} and space-dependent activity\cite{fernandez-rodriguez_feedback-controlled_2020}. 
The collective dynamics and emergent phenomena observed in active matter systems challenge traditional equilibrium concepts\cite{bischofberger_editorial_2023}.

Understanding the behavior of APs under confinement is a topic of increasing interest, particularly when considering their interactions with droplets and vesicles, which acts as a deformable medium with a well-defined equilibrium structure, and are commonly found in biological systems\cite{needleman_active_2017,aranson_bacterial_2022,fang_nonequilibrium_2019}.
Numerous works have been devoted to systems of APs within vesicles. 
Bacteria have been found to induce non-equilibrium shape fluctuations in giant vesicles\cite{takatori_active_2020}. 
Active filaments can both deform the vesicle shape and undergo collective motion while enclosed by the vesicle\cite{peterson_vesicle_2021}. 
Moreover, the shape of the vesicle can be greatly deformed by the presence of APs\cite{paoluzzi_shape_2016,li_shape_2019}, leading to a rich state diagram in terms of activity and AP concentration\cite{vutukuri_active_2020,iyer_dynamic_2023}. 
Recently, Quincke rollers have been found to spontaneously form vortices when confined by droplets and, moreover, can induce active fluctuations in the contour of 2D droplets\cite{kokot_spontaneous_2022}.
Previous research has shown that the interplay between APs and confinement can lead to intriguing effects on both the particles and the morphology of the confining medium\cite{diaz_emergent_2023}.

The complex behaviour of systems composed of APs confined by droplets motivates the use of  simulations to gain insight over their dynamic and steady-state properties.  
Simulations allow to study a wide range of activity regimes which can be expected to result in a rich  variety of droplet deformations. 

In this work we will perform a systematic study of the state behaviour of APs in phase-separated droplets -due to surface tension of a binary mixure (BM)- as a notable case of soft deformable media containing active inclusions. 
We aim to obtain the most relevant parameters that control both the placement of APs and the droplet morphology. 
A mesoscopic model capturing the responsiveness of the interface shape to the interaction with APs will allow to study the shape fluctuations in the presence of activity. 

\section{Model}

A mesoscopic hybrid model is used to study a system composed of $N_p$ APs with diameter $\sigma$ with translational degrees of freedom determined by $\rvec_i$ and orientational ones by $\phi_i$ , dispersed in a BM of two species A and B, defined by the differences of concentration of each species.
The order parameter of the BM $\psi(\rvec,t)=\phi_A-\phi_B$ characterises the local degree of demixing, with $\phi_A$ and $\phi_B$ being the local concentrations of species A and B, respectively. 
The conserved mean value of $\barpsi=\langle\psi \rangle$ specifies the global asymmetry of the BM.

The thermodynamic state of the system in the passive limit is specified by the total free energy, which has three terms, 
\begin{equation}
    F = 
    F_{BM} + F_{cpl}+F_{pp}
\end{equation}
respectively, free energy of the BM, coupling term and particle-particle interaction term. 
The free energy is assumed to be expressed in units of $k_BT$.

The BM free energy is a standard Ginzburg-Landau functional of $\psi(\rvec,t)$,
\begin{equation}
    F_{BM} = \int d\rvec \left[ 
    -\frac{1}{2} \tau \psi^2 +\frac{1}{4} u \psi^4
    +\frac{1}{2} D (\nabla \psi)^2
    \right],
    \label{eq:BM}
\end{equation}
where $\tau$ is related to the Flory-Huggins of the BM, $u$ specifies the amplitude of the concentration fluctuations and $D$ controls the width of the interface. 
In fact, the amplitude of the fluctuations can be analytically derived from the local terms in Eq. \ref{eq:BM} as $\psi_{eq}=\sqrt{\tau/u}$ which we refer to as the equilibrium values of the order parameter, that is, the values that $\psi$ takes in the bulk region when demixing takes place. 

The coupling free energy is\cite{pinna_modeling_2011,diaz_hybrid_2022}
\begin{equation}
    F_{cpl} = \sum_{i=1,N_p} c \int d\rvec \psi_c(|\rvec-\rvec_i|) \left[ \psi(\rvec)-\psi_0  \right]^2
    \label{eq:cpl}
\end{equation}
where $c$ specifies the scale of the particle-field interaction, $\psi_0$ specifies the selectivity of the BM towards the particle and $\psi_c$ is a tagged function that determines the size of the particle.  
We choose a functional form\cite{tanaka_simulation_2000}
\begin{equation}
    \psi_c(r<R)=\exp\left[1-\frac{1}{1-(r/R)^2} \right]  
\end{equation}
and $\psi_c(r\geq R)=0$ which has a compact form and vanishing derivative at the cutoff $r=R$. 

The particle-particle free energy term is
\begin{equation}
    F_{pp} = 
    \sum_{i\neq j} U(r_{ij})
\end{equation}
where the pairwise additive repulsive potential is 
$ U(r_{ij}) = U_0 exp(1-r_{ij}/\sigma)/(r_{ij}/\sigma) $
where $U_0$ specifies the energetic scale of the repulsive particle-particle interaction and $\sigma=2R$ is the diameter of the particle and also the cut-off of the interaction $U(r_{ij}>\sigma)=0$ where $r_{ij}$ is the distance between particle pairs.

We consider over-dumped diffusive dynamics both for the BM and the collection of APs (translational and rotational degrees of freedom). 
For the BM we consider Cahn-Hilliard-Cook\cite{cahn_free_1959,cahn_free_1959-1,cook_brownian_1970} dynamics governing the time evolution of the order parameter $\psi(\rvec,t)$ as 
\begin{equation}
    \frac{\partial \psi}{\partial t} = 
    M \nabla^2 \left( \frac{\delta F}{\delta \psi} \right)
    +\eta_{BM}(\rvec,t)
    \label{eq:Cahn}
\end{equation}
where $M$ is a mobility parameter that controls the diffusive time scale of the BM $t_{BM}\sim M^{-1}$. 
The functional derivative can be made explicit as $\delta F/\delta \psi=-\tau \psi+u\psi^3 -D\nabla^2\psi +2c\psi_c (\psi-\psi_0)$. 
Meanwhile, the random noise term satisfies the fluctuation-dissipation theorem as\cite{ball_spinodal_1990}
\begin{subequations}
    \begin{equation}
        \langle \eta(\rvec,t)\rangle  =0
    \end{equation}
    \begin{equation}
            \langle \eta(\rvec,t) \eta(\rvec',t')\rangle = -k_B T M \nabla^2 \delta(\rvec-\rvec')
\delta(t-t'). 
    \end{equation}
\end{subequations}

On the other hand, the collection of APs follow the active Brownian particle (ABP) model with additional forces arising from the coupling interaction with the BM, eq. \ref{eq:cpl}, $\fvec_i^{cpl}=-\nabla_i F_{cpl}$, on top of the particle-particle interaction $\fvec_i^{pp}=-\nabla_i F_{pp}$.
The total conservative force acting on particle $i$ is then $\fvec_i=\fvec_i^{cpl}+\fvec_i^{pp}$, leading to the diffusive dynamics
\begin{subequations}
\begin{equation}
    \frac{\partial \rvec_i}{\partial t} = 
    v_a \nhat_i + \fvec_i/\gamma_t + 
    \sqrt{2D_t} \xi_{t,i}(t)
    \label{eq:brown.t}
\end{equation}
\begin{equation}
    \frac{\partial \phi_i}{\partial t} = 
    \sqrt{2D_r} \xi_{r,i} (t)
    \label{eq:brown.r}
\end{equation}
\label{eq:brownian}
\end{subequations}  
where the fluctuation-dissipation for the translational degrees of freedom species that  
$\langle \xi_{t,i}(t) \rangle=0$ and $\langle \xi_{t,i}^{\alpha}(t)\xi_{t,j}^{\beta}(t') \rangle=\delta_{ij} \delta_{\alpha\beta} \delta(t-t')$
while for the rotational it holds that
$\langle \xi_{r}(t) \rangle=0$ and $\langle \xi_{r}(t)\xi_{r}(t') \rangle=\delta_{ij}\delta(t-t')$.

The Einstein relation applies for each diffusive constant $D_t=k_BT/\gamma_t$ and $D_r=k_BT/\gamma_r$, while  $D_r/D_t=3/\sigma^2$.
The active motion is characterised by the active velocity $v_a$ with a direction dictated by the unit vector $\nhat_i=(\cos\phi_i,\sin\phi_i)$.
A characteristic swimming time scale $t_{s}=\sigma/v_a$ and a rotational diffusive time scale  $t_{rot}=D_{r}^{-1}$ can be extracted and their ratio can characterise the persistence of the active motion given by the Péclet number $Pe=t_{rot}/t_{s}$. 

APs are capable of performing work given by $\varepsilon_a=\gamma_t v_a \sigma$, which can be compared with the other energetic scales of the model: 
thermal scale $k_BT$, particle-particle interaction strength $U_0$ and, most importantly, the coupling energy scale $\varepsilon_{cpl}=c\sigma^d \psi_{eq}^2$, which can be extracted from eq. \ref{eq:cpl}, with $d$ being the dimension of the system.

Throughout this work we select $\sigma$ as the unit of length, $t_{rot}$ as the unit of time and  $k_BT$ as the unit of energy. 
The order parameter $\psi$ will be expressed in units of its equilibrium values $\psi_{eq}$. 
We will restrict to two-dimensional systems $d=2$ with total system area $A=L_x~L_y$ and imposing periodic boundary conditions. 
Unless otherwise stated, all systems are initialised at $t=0$ with an circular droplet configuration with fixed mean value of the order parameter $\barpsi=-0.61$, leading to a positive droplet with $\psi=+1$ placed at the centre of the system with a radius $R_{drop}/L_x=1/4$ immersed in a matrix of $\psi=-1$. 
All APs are initially enclosed by the droplet, so that in the limit $Pe\to 0$ the initial configuration is also an equilibrium one.

We use a hybrid\cite{diaz_hybrid_2022} in-grid/out-of-grid numerical scheme where the continuous field $\psi$ is resolved using cell dynamic simulations\cite{oono_computationally_1987} in a lattice with spacing $\delta x$. 
On the other hand, the dynamics of the APs is resolved using standard forward Euler algorithm for Brownian dynamics. 
The presence of several time scales in the system (relaxation of the BM $t_{BM}$, rotational diffusive $t_{rot}$ and active $t_s$) requires that we consistently implement a discretisation scheme where the time step $\delta t$ is scaled with the fastest time scale in the system $\tilde{\delta t}=\delta t/ t_{faster}$ where $t_{faster}$ is the fastest (smallest) time scale present in the system.

\subsection{Observables}

To facilitate reproducibility we list here the quantities used to analyse the behaviour of the systems under study. 

The \textbf{mean square displacement} (MSD) is calculated as 
\begin{equation}
    MSD(t) = \frac{1}{N_p} \sum_i \left[ \rvec_i(t)-\rvec_i(0) \right]^2.
\end{equation}

The \textbf{average value} of $\psi$ inside each AP is calculated as 
\begin{equation}
    \barpsi_{AP}^i=\frac{4}{\pi\sigma^2}\int_{r<R}d\textbf{r} \psi(\rvec)
\end{equation}
which can be used to determine the overall average value of $\psi$ within the APs as $\barpsi_{AP}=\langle \barpsi_{AP}^i \rangle$. 
Furthermore, it can be used to determine the fraction $\Phi$ of APs such that $\barpsi_{AP}^i>0$.

We calculate each \textbf{pressure contribution} using standard expressions from the trace of the stress tensor  $p_{k}=Tr(\sigma_{\alpha\beta}^k)$ where $k$ is each of the pressure contributions and $\alpha\beta$ are the  Cartesian components of the tensor. 
For the BM pressure the contributions arise from the interfaces of the BM\cite{ohta_anomalous_1993}
\begin{equation}
    \sigma_{\alpha\beta}^{BM}=
    -\frac{D}{A} \int d\rvec \partial_{\alpha} \psi \partial_{\beta} \psi  
\end{equation}
where $\partial_{\alpha}$ is the spatial derivative in the $\alpha$ Cartesian coordinate. 

The particle-particle interacting pressure $p_{pp}$ arises from the repulsive forces between APs. 
We use the virial expression 
\begin{equation}
    \sigma_{\alpha\beta}^{pp}=
    \frac{1}{2A} \sum_{ij} f_{ij}^{\alpha}x_{ij}^{\beta}
\end{equation}
where $i,j$ are particle pairs, $f_{ij}^{\alpha}$ is the repulsive force and $x_{ij}$  is the distance between particles. 

The active pressure is calculated as\cite{solon_pressure_2015,speck_ideal_2016}
\begin{equation}
    p_{a} = 
    \frac{v_a \gamma_t}{2V} \sum_i \langle \nhat_i \cdot \rvec_i \rangle
\end{equation}

For the coupling pressure we consider that AP $i$ interacts with the embedding BM through the order parameter field $\psi(\rvec,t)$. 
For this reason, we use a virial-like pressure calculation where particle $i$ in continuous space $\rvec_i$ is considered to interact with fictitious particle $j$ corresponding to node point $\rvec_j$ corresponding to $\psi(\rvec_j)$. 
The coupling force contribution $ f_{ij}^{cpl,\alpha}$
\begin{equation}
    \sigma_{\alpha\beta}^{cpl}=
    \frac{1}{2V} \sum_{ij} f_{ij}^{cpl,\alpha}x_{ij}^{\beta}
\end{equation}
where $f_{ij}^{\alpha}$ is the $\alpha$ component of the force contribution arising from the interaction between particle $i$ and node point $\rvec_j$. 
This force arises from the gradient of the coupling free energy in Eq. 1. 
On the other hand, the distance $x^{\beta}_{ij}$ is the $\beta$ component of the distance between particle $i$ position (continuous) and node point $\rvec_j$, $x_{ij}^{\beta}=x_i^{\beta}-x_j^{\beta}$. 
Conceptually, we can understand that particle $i$ interacts with he collection of molecules at node $\rvec_j$. 

\textbf{Domain analysis} tools can be used to determine the interface points for a given field state $\psi$. 
Using cluster analysis each droplet can be studied separately. 
For a given droplet the interface points $\rvec_{\alpha}$ can be used to calculate the centroid $\textbf{R}_c$ and the distances to the centre of the droplet $\Delta \rvec_{\alpha}=\rvec_{\alpha}-\textbf{R}_c$ (see scheme in Fig. \ref{fig:scheme}). 
The droplet radius can be calculated as $R_{drop}=\langle |\Delta \rvec_{\alpha}| \rangle$ and the standard deviation of the droplet radius can be calculated from the fluctuations of the droplet radius $h_{\alpha} = \rvec_{\alpha}-R_{drop}$, as 
\begin{equation}
    \sigma_R^2 = 
    \langle \left( |\Delta\rvec_{\alpha}| -  R_{drop} \right)^2 \rangle .
\end{equation}

Additionally, the droplet radius can be tracked over time $R(t)$ and the steady state droplet size $R^*$ can be determined.

\section{Results}

Throughout this work we will use parameters $\tau=0.35$, $u=0.5$, $D=0.25$, $M=0.25$ for the BM. 
For the APs we consider $c=1$, $\psi_0=+1$. 
The discretisation scheme is $\delta x=0.296$ and $\tilde{\delta t}=0.01225$

In this work we will first study the dilute regime in order to gain insight on the placement and dynamics of APs in the near-single-particle limit, where droplet deformations due to APs can be expected to be minimal. 
In this step key dimensionless parameters will be extracted. 
Secondly, we will explore the state diagram of the coupled AP/BM system for various degrees of activity and concentration.
Finally, we will concentrate on the shape fluctuations due to activity for various droplet sizes. 

\subsection{Dilute regime}

\begin{figure}
    \centering
    \includegraphics[width=0.99\textwidth]{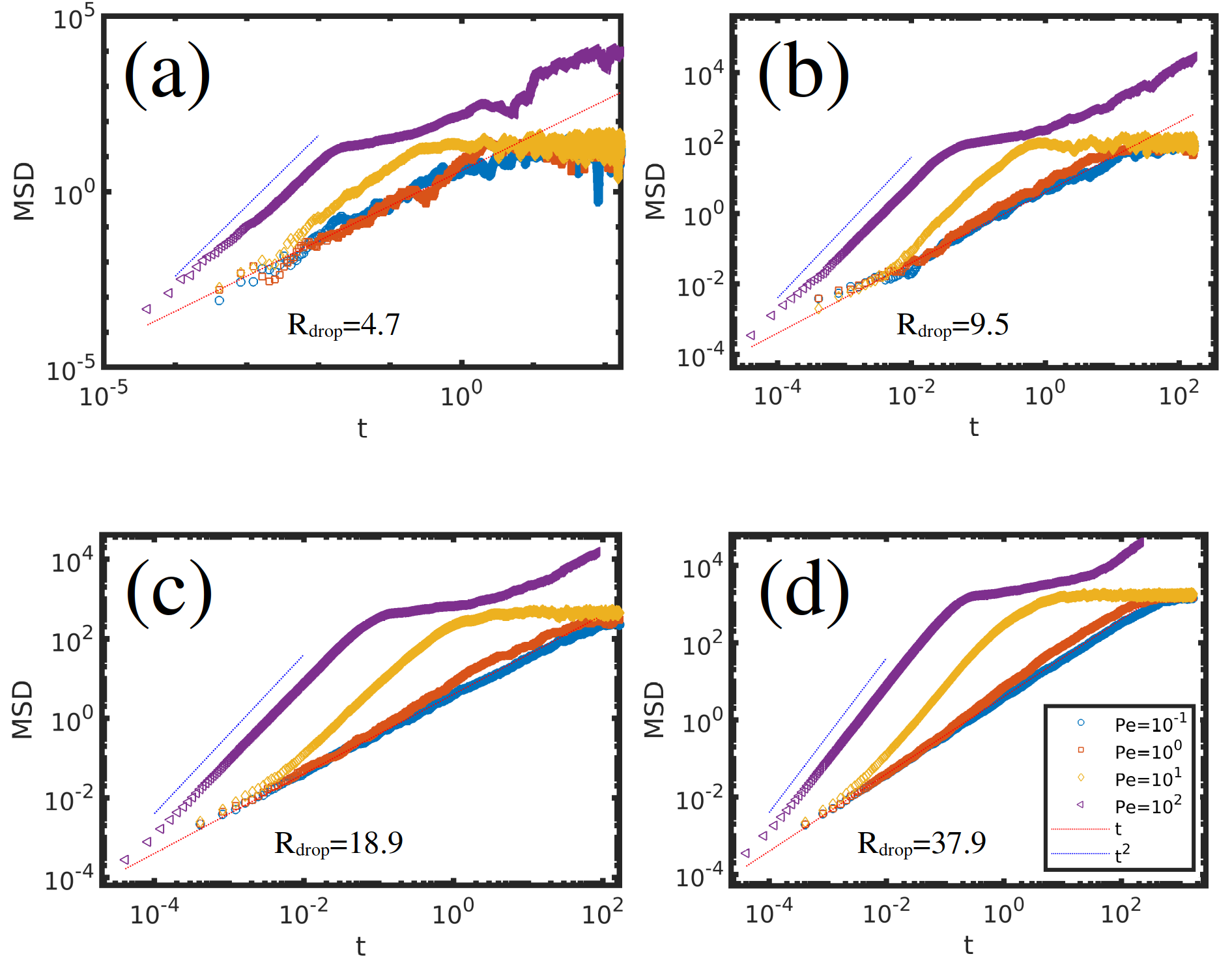}
    \caption{
        Mean Square Displacement in the dilute regime $\phi_p=0.01$ for various droplet sizes with 
        (a) $R_{drop}=4.7$,
        (b) $R_{drop}=9.5$,
        (c) $R_{drop}=18.9$ and 
        (d) $R_{drop}=37.9$. 
        See movies S1, S2, S3 and S4 for the corresponding time evolution for $R_{drop}=18.9$. 
    }
    \label{fig:msd}
\end{figure}
We first consider a highly dilute concentration of APs $\phi_p=0.01$ with four representative activities quantified by $Pe$. 
Fig. \ref{fig:msd} shows the effect on confinement on the MSD depending on the activity and the droplet size. 
For all values of $R_{drop}$, for small activity $Pe=0.1$ we observe the expected diffusive behavior for a free particle $MSD=4D_t t$ until APs have explored a length scale comparable with the droplet size, when the droplet acts as a confining medium (see movies S1 and S2). 
For  moderately active APs $Pe=10$ the MSD is that of ABPs\cite{howse_self-motile_2007} with a ballistic diffusion $MSD\propto t^2$ for intermediate times.
However, activity is not large enough to deform the droplet and therefore it continues to act as a confining environment limiting the growth in the MSD (see movie S3). 
The role of the droplet size is clear as for increasingly larger values of $R_{drop}$ the upper bound of the permitted $MSD(t\to\infty)$ increases accordingly.

\begin{figure}
    \centering
    \includegraphics[width=0.5\linewidth]{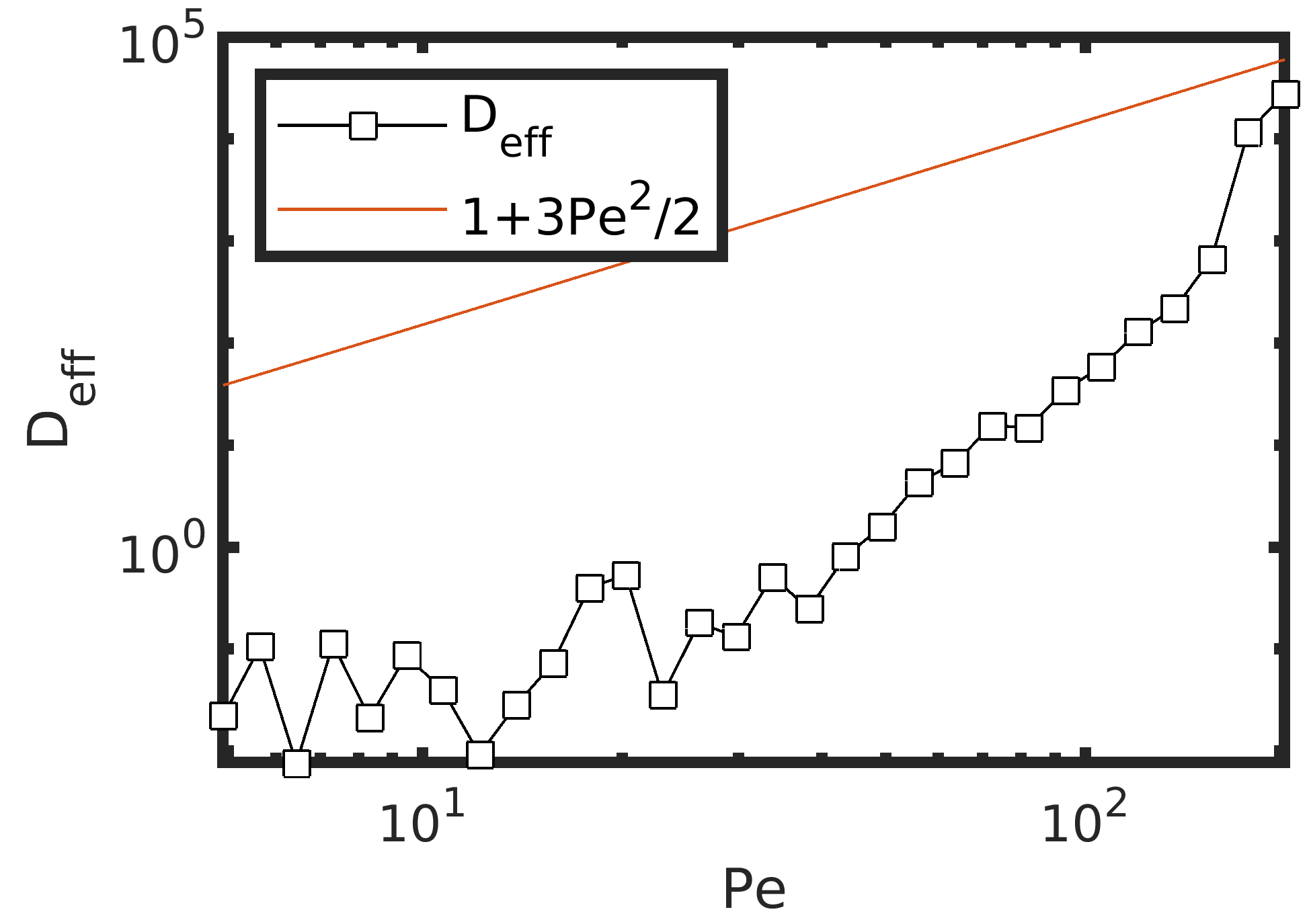}
    \caption{
        Long-time effective diffusion in function of $Pe$, calculated from the limit $MSD(t\to\infty)$. 
        Solid points are the numerical calculation while the ABP analytical expression is shown as a line 
    }
    \label{fig:msd.Deff}
\end{figure}

Finally, for highly activity $Pe=100$ the MSD exhibits a transient ballistic regime $MSD \propto t^2$ for intermediate times, followed by a long-time diffusive behaviour $MSD\sim t$. 
For such high activity regimes APs can break the confinement induced by the droplet leading to their escape into the bulk incompatible phase (see movie S4). 
This occurs because the active energy per particle $\varepsilon_a$ is comparable with the coupling one associated with the interaction with the medium $\varepsilon_{cpl}$. 
The long-time effective diffusivity of the APs $D_{eff}=MSD(t\to \infty)/4t$ can be compared with the expected analytical expression for a suspension of APs  $D_{eff}=D_t(1+3/2 Pe^2)$. 
Fig. \ref{fig:msd.Deff} shows the $D_{eff}$ in terms of $Pe$ to be several orders of magnitude below the nominal value (solid line), which indicates that APs slow down in the presence of repulsive forces when they enter the gray phase, due to the coupling forces arising from the particle-medium surface tension. 

\begin{figure}
    \centering
    \includegraphics[width=1.0\linewidth]{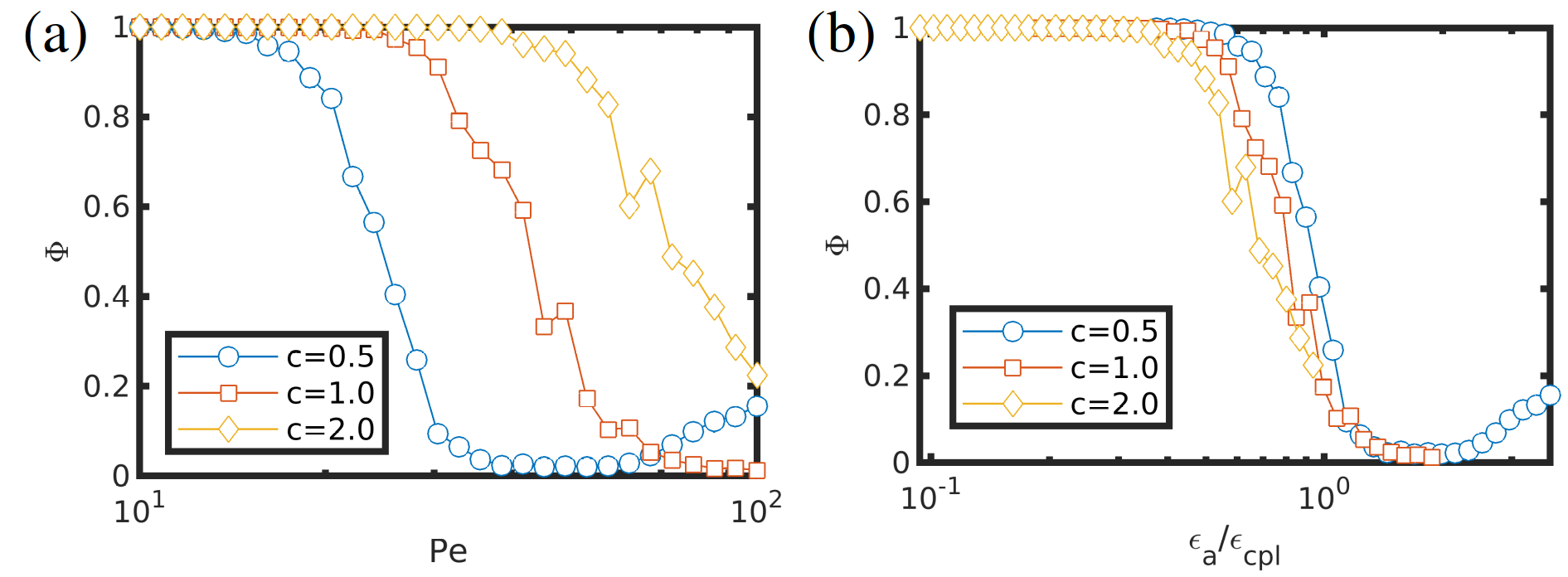}
    \caption{
    Effect of confinement on  ABPs. 
    The fraction of ABPs in the white phase $\Phi$ is shown in terms of $Pe$ in (a) while the active energy is scaled with respect to the coupling energy scale is shown in (b). 
    }
    \label{fig:fraction}
\end{figure}

In order to better quantify the ability of activity to allow APs to escape soft confinement, we explore $Pe$ for a $R_{drop}=9.5$ in the same dilute regime $\phi_p=0.01$. 
The fraction of APs within the positive (white) phase $\Phi$ is shown in Fig. \ref{fig:fraction} (a) for various values of $c$, which quantify the strength of the AP-BM interaction and can be related to the surface tension of the AP with respect to the incompatible negative (gray) phase. 
As expected, smaller values of $c$ lead to smaller critical $Pe$ value upon which $\Phi$ decreases, indicating a reduction in the activity needed to allow APs to escape. 
The coupling energy parameter $\varepsilon_{cpl}$ can be used to estimate the energetic cost associated to the exposure of the AP into the gray phase (\textit{i.e.} surface tension). 
Fig. \ref{fig:fraction} (b) shows that the various curves of $\Phi$ can be approximately collapsed into a single one when rescaling the $x$ axis as the ratio between the active energy $\varepsilon_a$ and the coupling one. 
When this ratio is comparable to one, APs can overcome the energetic barrier associated to the interface and jump into the bulk gray phase. 
As Fig. \ref{fig:msd} shows, once APs enter the gray phase they do not behave as confined free particles, but they exhibit a reduced diffusivity due to the coupling forces arising from the incompatibility of the AP in the surrounding medium. 
Notably, for high $\varepsilon_a/\varepsilon_{cpl}\gg 1$ the fraction $\Phi$ begins to grow, indicating that APs are homogeneously distributed in space, as can be confirmed in the limit of $Pe\to\infty$ in Fig.~\ref{fig:fraction.Peinf} as $\Phi\to \pi R_{drop}^2/V\approx 0.196$. 

In the dilute regime activity can drive APs away from their equilibrium configuration in the solvable phase and into the insolvable phase, as  APs can perform work necessary to overcome the energetic barriers associated with the surface tension of the APs when exposed to the gray phase. 
However, at higher concentrations we can expect that the active energy associated to the active motion can be comparable with the BM surface tension itself, which motivates a detailed study of the role of the AP concentration. 

\subsection{Droplet behaviour in terms of AP concentration and activity}

\begin{figure}
    \centering
    \includegraphics[width=0.5\linewidth]{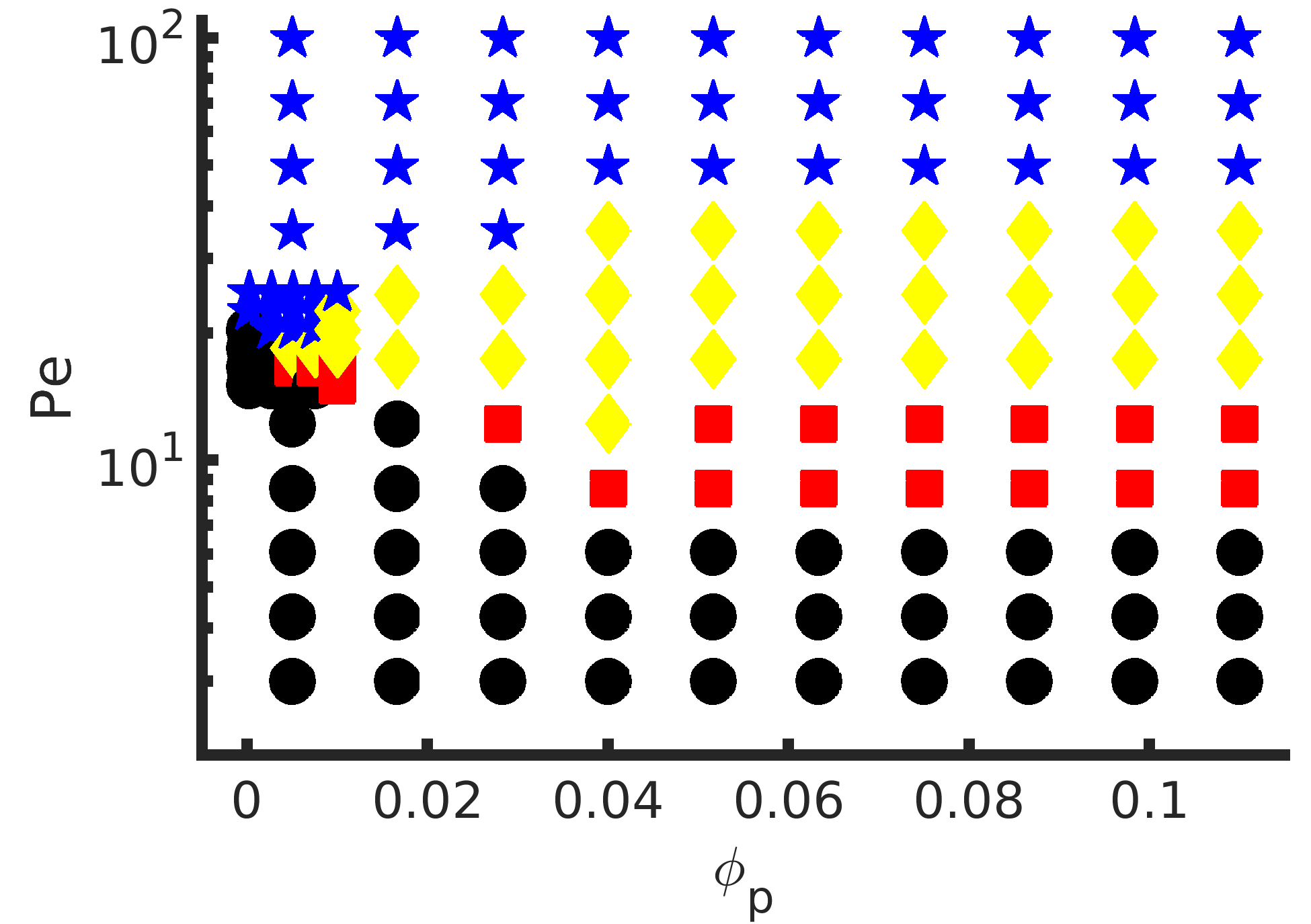}
    \caption{
    State diagram of a droplet with size $R_{drop}=18.9$ in the presence of a concentration $\phi_p$ of APs with activity given by $Pe$. 
    Black circles correspond to an isotropic droplet morphology enclosing APs. 
    Red squares correspond to a broken droplet morphology. 
    Yellow diamonds indicate a broken droplet morphology with homogeneous spatial distribution of APs.
    Blue stars indicate droplet morphology with escaped APs. 
    }
    \label{fig:phd.L256}
\end{figure}

\begin{figure}
    \centering
    \includegraphics[width=0.5\linewidth]{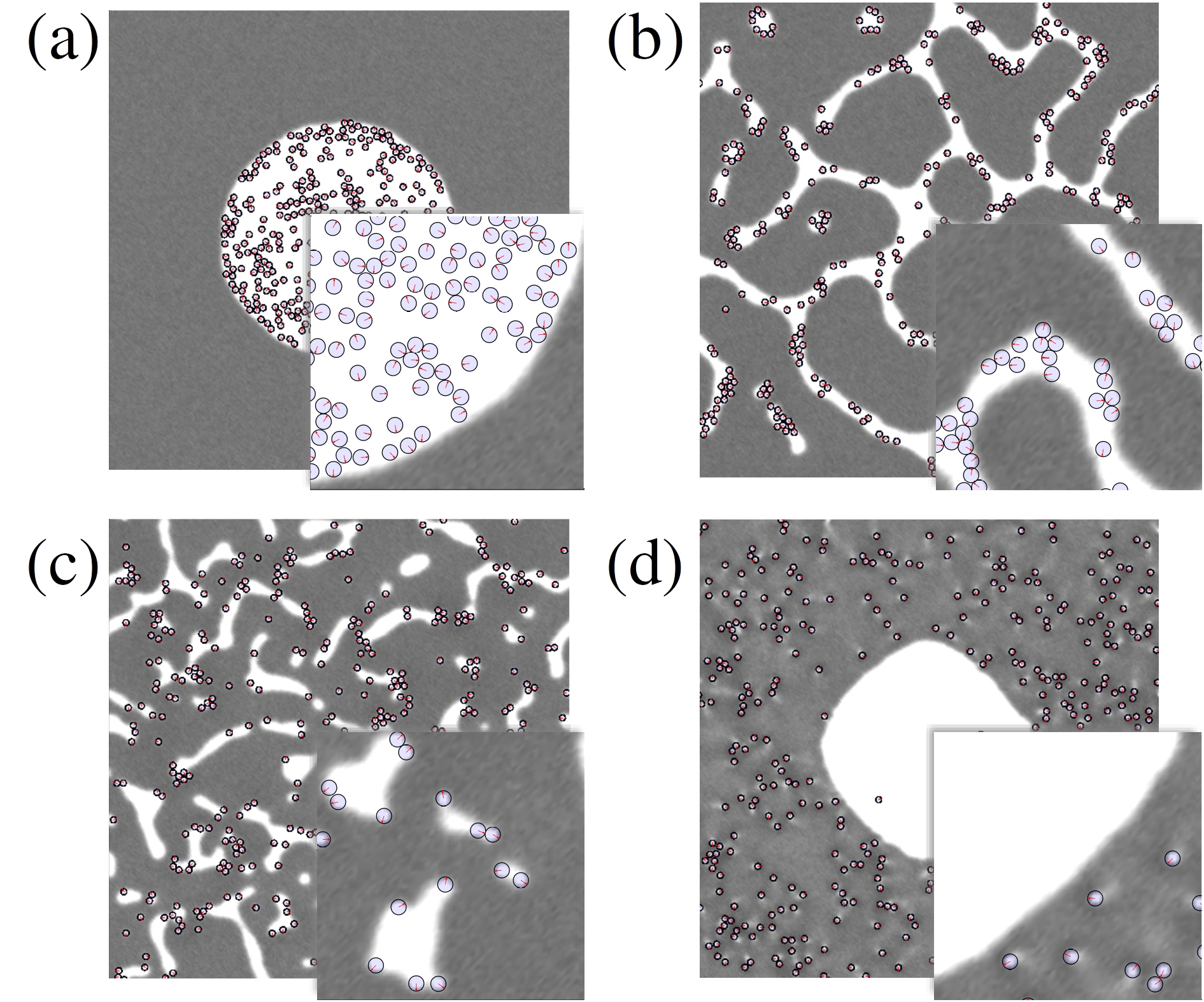}
    \caption{Representative snapshots of a droplet with size $R_{drop}=18.9$ for $\phi_p=0.05$ and increasing values of activity
    (a) $Pe=3$, 
    (b) $Pe=12$, 
    (c) $Pe=25$ and  
    (d) $Pe=100$, 
    corresponding to the state diagram in Fig. \ref{fig:phd.L256}. 
    The red arrow for each AP indicate the direction of self-propulsion. 
    }
    \label{fig:phd.L256.snaps}
\end{figure}

\begin{figure}
    \centering
    \includegraphics[width=1.0\linewidth]{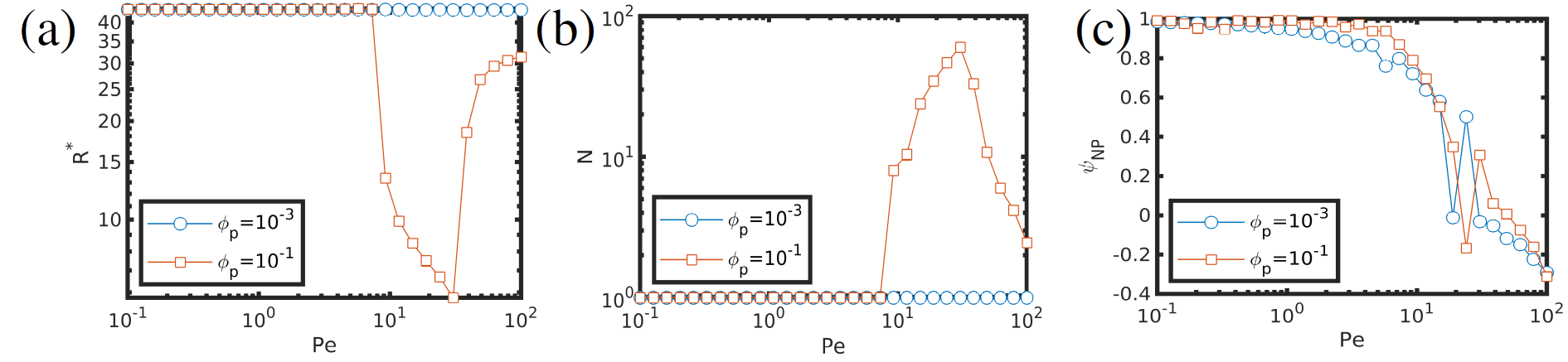}
    \caption{
        Curves of observables for $R_{drop}=18.9$ used to determine the state diagram in Fig. \ref{fig:phd.L256}. 
        Two representative AP concentrations are shown $\phi_p=10^{-3}$ and $10^{-1}$. 
        In (a) the characteristic droplet size $R^*$ is shown in terms of $Pe$. 
        In (b) the total number of droplets in the system is displayed $N$. 
        In (c) the average value of $\psi$ inside the particle is shown. 
    }
    \label{fig:curves.L256}
\end{figure}

\begin{figure}
    \centering
    \includegraphics[width=0.5\linewidth]{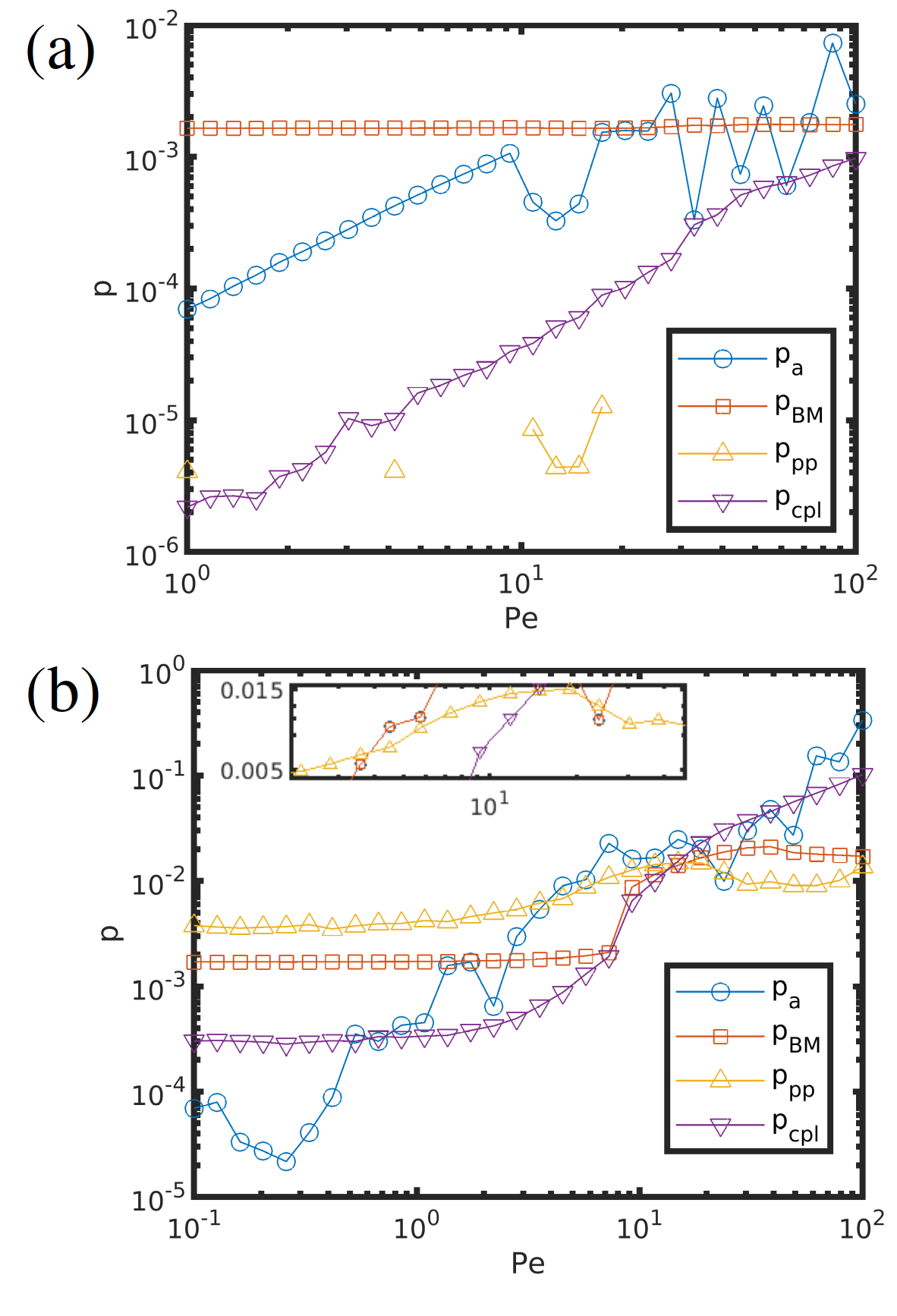}
    \caption{
        Pressure profiles for  $R_{drop}=18.9$ in function of $Pe$ for(a) $\phi_p=10^{-3}$ and (b) $\phi_p=0.1$.  
        The components are
        active pressure $p_a$, 
        BM pressure $p_{BM}$, 
        interaction particle-particle $p_{pp}$ and 
        interaction particle-field $p_{cpl}$. 
        The $p_{pp}$ pressure in (a) rarely takes nonzero values due to the dilute regime, leading to infrequent events of particle-particle collisions.
    }
    \label{fig:pressure.L256}
\end{figure}

We consider a droplet with size $R_{drop}=18.9$ which contains a concentration $\phi_p$ of APs with activity given by $Pe$. 
Considering both the AP localisation and the droplet morphology, we can identify several regimes which we classify in the state diagram in Fig.~\ref{fig:phd.L256} based on visual inspection as shown in Fig.~\ref{fig:phd.L256.snaps} and the observables shown in Fig.~\ref{fig:curves.L256} for two representative AP concentrations (see Fig.~\ref{fig:phd.L256.colormaps} for the full colourmaps). 
In regime (I), for small activity, APs are completely dispersed within their solvable phase, marked as black circles in Fig. \ref{fig:phd.L256}. 
A representative snapshot is shown in Fig.~\ref{fig:phd.L256.snaps} (a). 
In this regime the droplet morphology remains unperturbed by the activity with $R^*=18.9$ for $Pe<10$ in Fig.~\ref{fig:curves.L256} (a) and $N=1$ in (b).
As we will see, this does not exclude the emergence of activity-induced fluctuations in the droplet interface contour, which will be studied in detail in the next section, but only reflects that the morphology of the droplet is preserved while all APs are enclosed by the droplet. 
Furthermore, the overall shape of the droplet can be deformed leading to increases in the asphericity, shown in Fig. \ref{fig:asphere}
for $\phi_p=0.1$ and $Pe\sim 7$, while a single droplet is present in the system. 
We note that APs can accumulate at the droplet boundaries due to the persistence of their motion, which is commonly observed in APs under confinement\cite{li_accumulation_2009,palacios_guided_2021},
leading to increases in the 
particle-BM coupling pressure and  the particle-particle pressure in Fig. \ref{fig:pressure.L256}, while the BM pressure remains independent of $Pe$ for small activity. 

\begin{figure}
    \centering
    \includegraphics[width=1\linewidth]{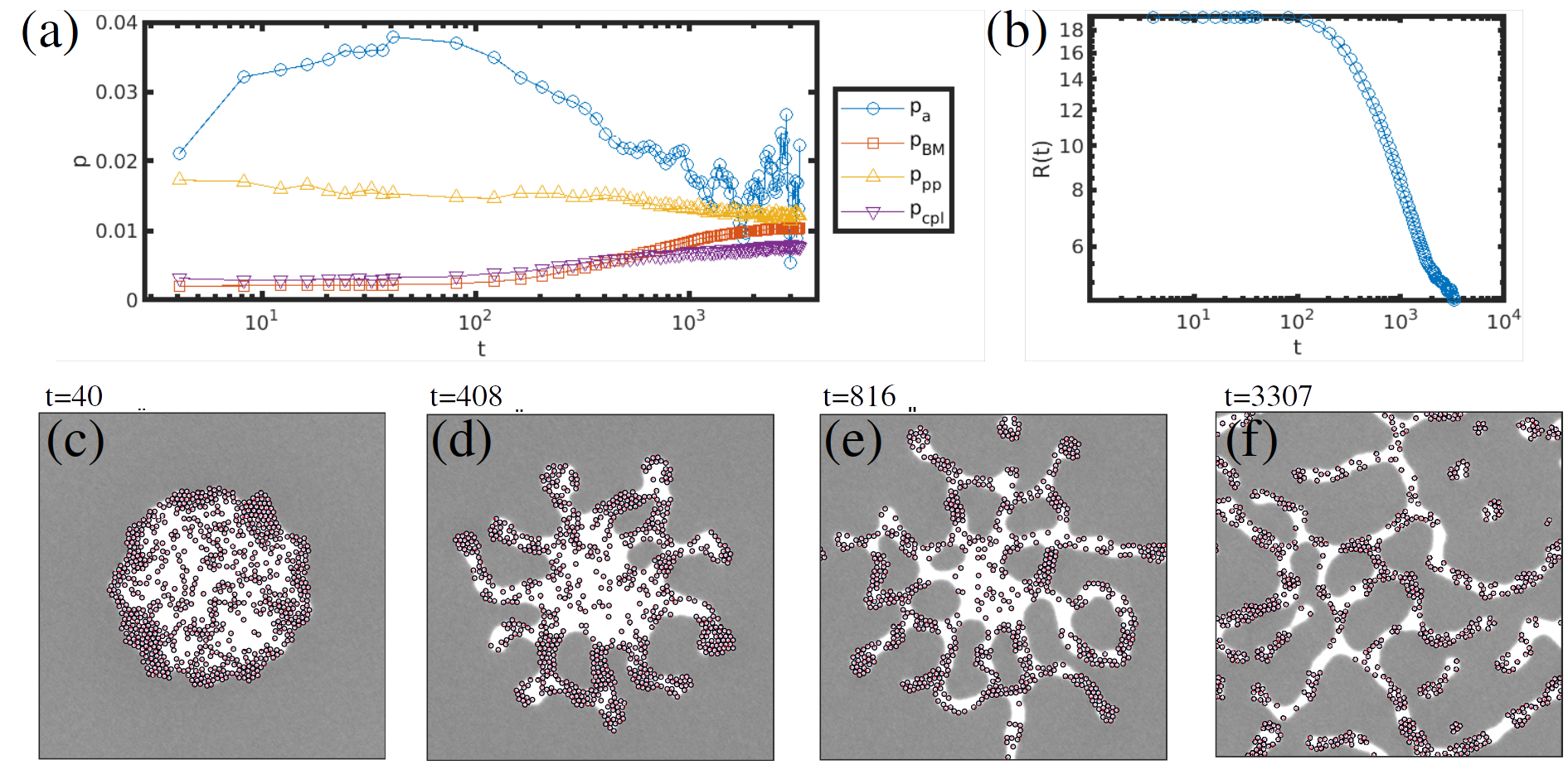}
    \caption{
        Droplet break with $R_{drop}=18.9$ for a system with $\phi_p=0.1$ and $Pe=10$ -regime II in Fig.\ref{fig:phd.L256}. 
        In (a) the components of the pressure are shown over time for 
        $p_a$ active pressure,
        $p_{BM}$ the BM,
        $p_{cpl}$ the coupling  and
        $p_{pp}$ the particle-particle one. 
        In (b) the characteristic droplet size is shown over time. 
        In (c), (d), (e) and (f) representative snapshots are shown    
        See Movie S5 for the full time evolution.
    }
    \label{fig:break.L256}
\end{figure}

For moderate to high concentrations we identify a different regime (II) where APs are capable of significantly disturbing the droplet morphology, marked as red squares in Fig.~\ref{fig:phd.L256}. 
In this regime, APs can perform active work against the droplet interfaces, as shown in Fig.~\ref{fig:phd.L256.snaps}(b), leading to a significant decrease in the steady-state droplet size $R^*$ and also the creation of smaller droplets, as shown in Fig.~\ref{fig:curves.L256} (a) and (b), respectively. 
We note that a critical concentration is required in order to trigger this transition $\phi_p>\phi_p^*\approx 2\cdot 10^{-3}$, which will be discussed later in this section. 
A closer look at the droplet breakage mechanism in Fig.~\ref{fig:break.L256} for $Pe=10$ and $\phi_p=0.1$ allows to identify an initial stage where APs accumulate at the droplet interface, where they exert active pressure on the soft-confining walls of the droplet. 
As the APs begin to form poles, they are capable of penetrating the interface walls pushing into the gray domains, with an increase in the BM pressure contributions $p_{BM}$ as the droplet morphology is disturbed, and an increase in the coupling pressure $p_{cpl}$ as more APs are exposed into the insolvable gray phase.  

In Fig.~\ref{fig:phd.L256}, for $\phi_p>\phi_p^*\approx 2\cdot10^{-3}$ but for higher activity, we identify a secondary transition leading to regime (III), marked as yellow diamonds. 
In this regime APs can not only render the droplet morphology unstable, but lead to a much more homogeneous distribution of APs within the system, as shown in Fig.~\ref{fig:phd.L256.snaps}(c)  where the white domains are considerably smaller (see reduction in $R^*$ in Fig.~\ref{fig:curves.L256}(a)) and APs can be found both in the white and gray domains. 
Another notable difference between regimes  II and III is the inversion of curvature: 
while the white phase is the minority one in the system, $\barpsi<0$, active pressure in regime II can stabilise gray domains in a white matrix which again changes to white domains in a gray matrix in regime III. 
This II-III transition is reminiscent of the one shown in the dilute regime in Fig.~\ref{fig:fraction} where APs are capable of performing enough work to overcome the energetic barrier associated with the interface $\varepsilon_a>\varepsilon_{cpl}$. 
However, this is different from regime II where APs could only overcome such barrier as clusters due to collective behaviour, which explains the formation of poles.

\begin{figure}
    \centering
    \includegraphics[width=1\linewidth]{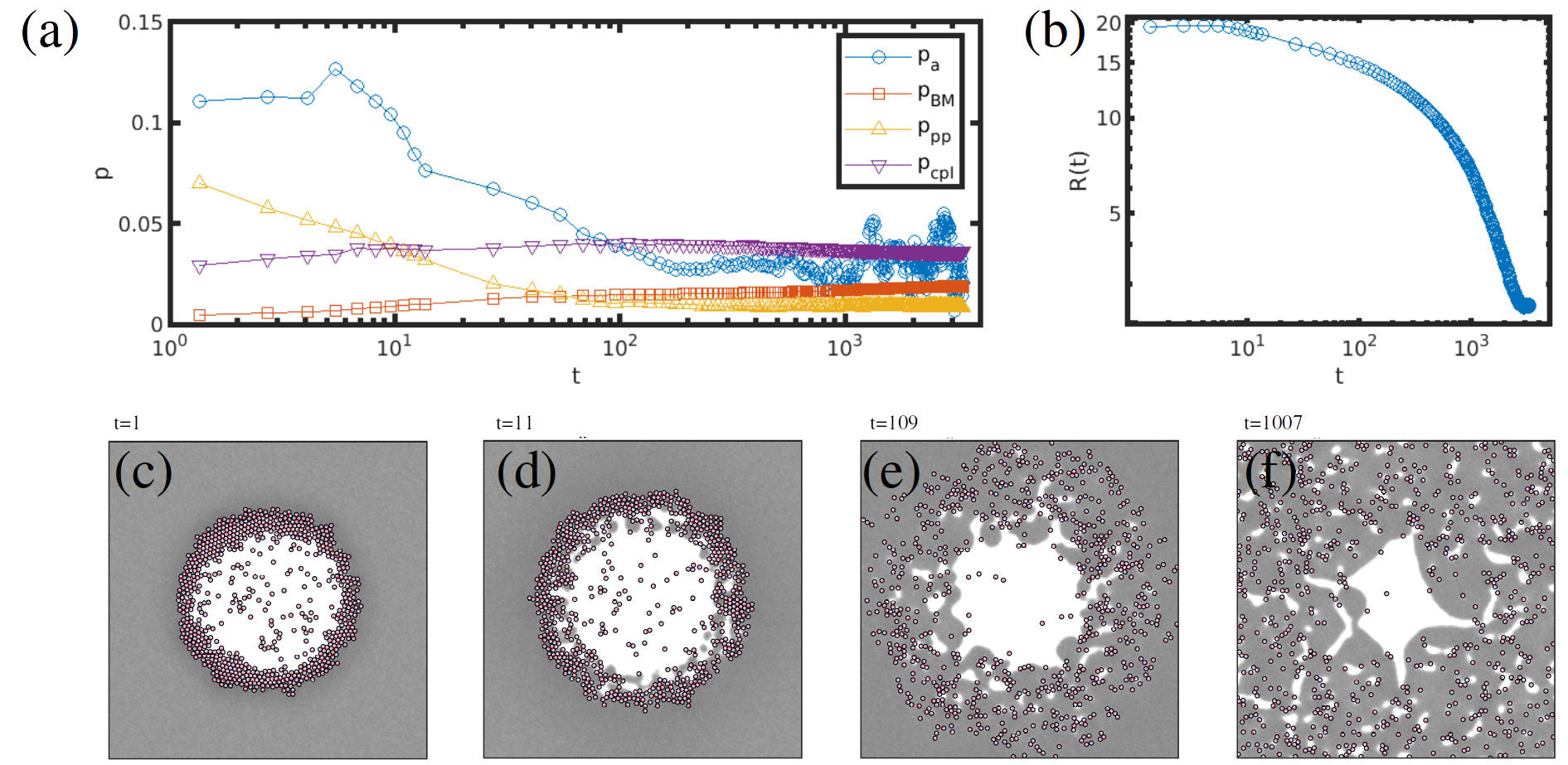}
    \caption{
        Droplet break with $R_{drop}=18.9$ for a system with $\phi_p=0.1$ and $Pe=30$ -regime III in Fig. \ref{fig:phd.L256}. 
        In (a) the components of the pressure are shown over time. 
        In (b) the characteristic droplet size is shown over time. 
        In (c), (d), (e) and (f) representative snapshots are shown.
        See Movie S6 for the full time evolution. 
    }
    \label{fig:break.L256.Pe30}
\end{figure}

In order to gain insight on the mechanism of droplet break within regime III, we consider a droplet with $R_{drop}=18.9$ enclosing $\phi_p=0.1$ APs with activity given by $Pe=30$. 
In Fig. \ref{fig:break.L256.Pe30} (a) we can observe the pressure components over time, which can  be contrasted with the behaviour shown in Fig.~\ref{fig:break.L256} (a) for regime II: 
it is clear that in regime III $p_{pp}$ decreases considerably faster as APs can escape the droplet boundaries without the need to form poles (see snapshot in \ref{fig:break.L256.Pe30}(d)).
In this sense, it can be argued that the escape mechanism for regime III occurs at the single-particle level, while for regime II it is a collective mechanism involving an intermediate step of clusterisation. 

A more quantitative way to differentiate between regime II and III is to observe the dependence of $p_{pp}$ with $Pe$ in Fig. \ref{fig:pressure.L256} (b) for $\phi_p=0.1$.
A sharp decrease in $p_{pp}$ can be observed for $Pe\leq 14.8$ 
( see inset in Fig. \ref{fig:pressure.L256} (b))
indicating that particle-particle collisions become less likely, resulting from the ability of APs to escape confinement as $\varepsilon_a \gg \varepsilon_{cpl}$. 
As APs have more available space to explore, the particle-particle collisions become less relevant, while in regime II APs are still highly confined by the BM, even if they have the ability to deform the droplet morphology.

Finally, for high activity we identify a regime (IV) which corresponds to the inversion of the AP location, from the solvable white phase into the insolvable gray one, and marked as blue stars in Fig~\ref{fig:phd.L256}.  
This can be quantified by the average value of $\psi$ around the particle $\psi_{NP}<0$ in Fig. \ref{fig:curves.L256} (b). 
Additionally, the BM recovers its isotropic droplet morphology, as shown in the snapshot in Fig.~\ref{fig:phd.L256.snaps} (d).  
In the dilute limit $\phi_p\ll \phi_p^*\approx 2\cdot10^{-3}$ this is simply the consequence of the APs overcoming the energetic barriers associated with the interface due to activity $\varepsilon_a>\varepsilon_{cpl}$ and accumulating at the gray phase, as discussing in detail in the previous section. 
However, for higher concentrations $\phi_p>\phi_p^*$ a higher $Pe$ is required in order to recover the droplet morphology.

\begin{figure}
    \centering
    \includegraphics[width=0.5\linewidth]{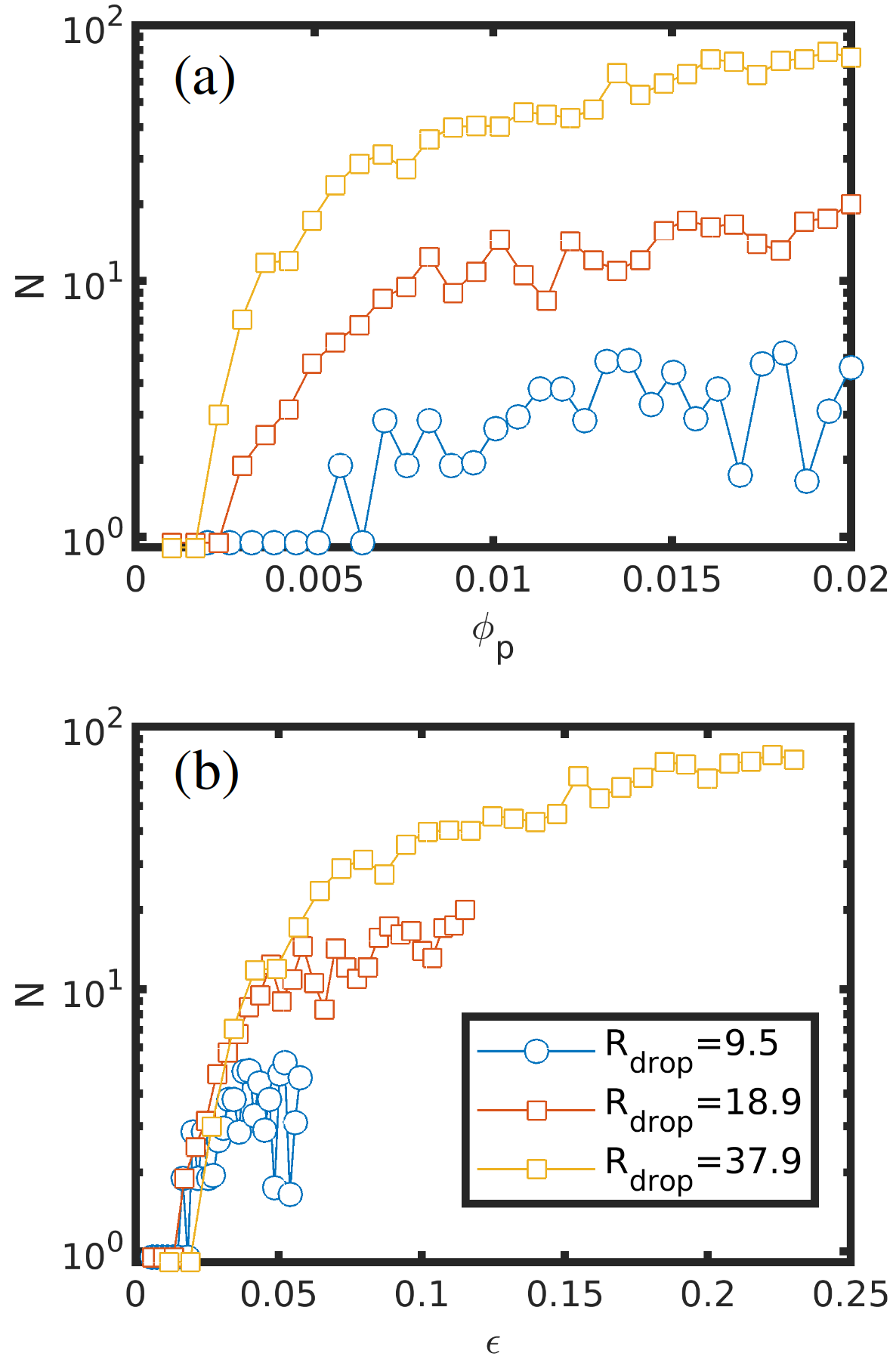}
    \caption{
        Effect of AP concentration $\phi_p$ on the droplet morphology characterised by the number of droplets $N$ in terms of the concentration  in (a) and the reduced energetic ratio $\varepsilon$ ( see Eq.~\ref{eq:epsilon}).  
    }
    \label{fig:phip}
\end{figure}

In Fig. \ref{fig:phd.L256} we have identified a critical point $\phi_p^*\approx 2\cdot 10^{-3}$ and $Pe^*\approx 19$, below which we observe a I-IV phase transition when increasing activity, while for $\phi_p>\phi_p^*$ it is possible to deform the droplet morphology into smaller droplets and the system can undergo three transitions I-II-III-IV.
By exploring the concentration of APs for a fixed $Pe=Pe^*$, in Fig.~\ref{fig:phip} we can identify the onset of the I-II transition. 
In (a) we can identify the departure from regime I by analysing the total number of BM droplets in the system, where $N>1$ marks the formation of additional droplets different from the main one, due to the presence of APs. 
It is clear that as $R_{drop}$ increases, the critical value of $\phi_p^*$ decreases. 
In order to gain insight on this dependency, an energy balance can be performed: 
on the one hand we can consider that total energy associated with the ability of APs to perform active work, estimated as $E_a=N_p\varepsilon_a$, while the total energy associated with the surface tension of the droplet interface is $E_{\gamma}=2\pi R_{drop} \gamma_{AB}$. 
A dimensionless parameter can be defined as
\begin{equation}
    \varepsilon = 
    \frac{E_a}{E_{\gamma}} = 
    \phi_p \frac{2k_BT Pe}{\pi^2 \gamma_{AB}} \frac{V}{R_{drop}\sigma^2}
    \label{eq:epsilon}
\end{equation}
which is shown in the horizontal axis in Fig. \ref{fig:phip}(b), leading to the collapse of the $N$ curve into a single one, suggesting that a critical value of $\varepsilon^*$ can be identified. 
This parameter can be understood as the competition between the total active energy in the system and the total energy associated with the interfaces. 
In other words, the total energy input that is capable of driving the droplet morphology away its equilibrium configuration. 
This energy ratio can be compared to the one shown in Fig. \ref{fig:fraction} in the dilute regime $\phi_p\ll \phi_p^*$, where the main energetic comparison was between the single-particle active work $\varepsilon_a$ and the interaction energy with the gray phase $\varepsilon_{cpl}$, related to the surface tension of APs with the insolvable phase. 

\subsection{Shape fluctuations}

In Fig.~\ref{fig:phd.L256} regime I has been identified as formed by a single droplet enclosing APs. 
Despite the morphology of the BM being that of an approximately isotropic droplet, fluctuations in the circular shape of the droplet can occur due to activity, as suggested by the coupling pressure profiles in Fig.~\ref{fig:pressure.L256} which increases with $Pe$ within regime I, even though the BM pressure remains flat. 
This can be attributed to the collisions and accumulation of APs at interfaces while acting as a soft confining medium.

\begin{figure}
    \centering
    \includegraphics[width=1\linewidth]{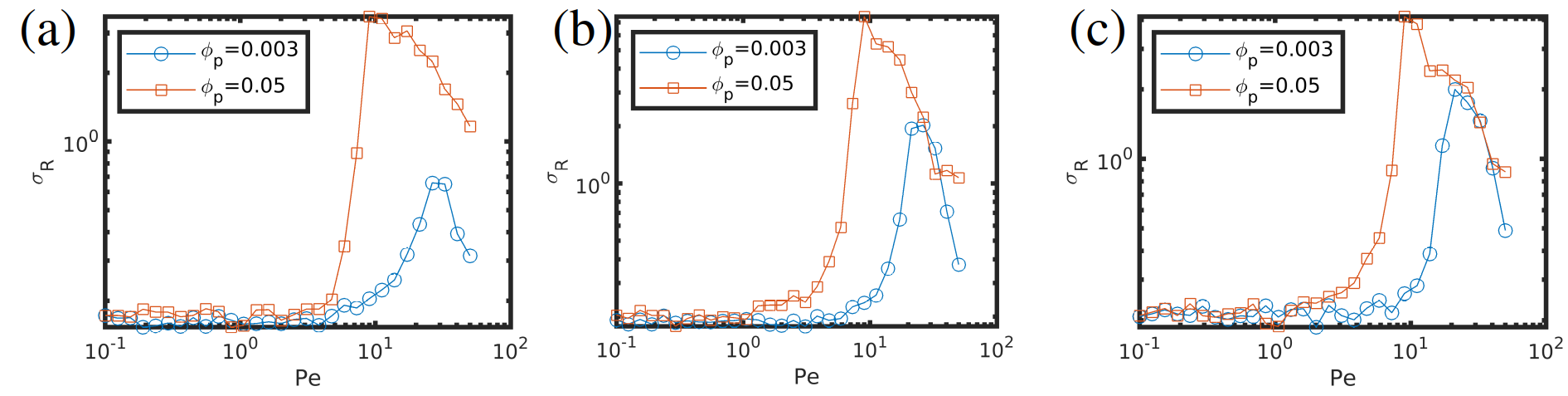}
    \caption{
        Fluctuations in the droplet size $\sigma_R$ for different droplet sizes in terms of activity $Pe$ for two AP concentrations: dilute with $\phi_p=0.003$ and concentrated with $\phi_p=0.05$. 
        Droplet sizes are $R_{drop}=9.5$ in (a), $18.9$ in (b) and $37.9$ in (c). 
    }
    \label{fig:fluctuations}
\end{figure}

Fig.~\ref{fig:fluctuations} shows the fluctuations in the droplet size $\sigma_R$ in terms of $Pe$ for two AP concentrations: 
dilute similar to the critical point with $\phi_p=0.003\sim \phi_p^*$ and above $\phi_p^*$ with $\phi_p=0.05$. 
Additionally, we consider three initial droplet sizes $R_{drop}$ to assess the role of the droplet dimension with fixed $\phi_p$. 
For both concentrations, in the small $Pe$ regime we recover passive fluctuations  $\sigma_R\sim 0.2$ due to the thermal fluctuation of the droplet interface. 
As activity increases, APs exert pressure on the interface leading to enhanced fluctuations $\sigma_R$. 
In the dilute regime $\phi_p=0.003$ this enhancement is moderate, but the droplet morphology is preserved across the range of $Pe$ values with regime I for $Pe<Pe^*\sim 19$ -where $\sigma_R$ grows with $Pe$- and regime IV for $Pe>Pe^*$ -where $\sigma_R$ decreases with $Pe$-.
For higher concentrations $\phi_p=0.05>\phi_p^*$ fluctuations grow rapidly with activity, but the transition into regime II occurs for much smaller $Pe$, indicating the departure from the single-droplet morphology into smaller droplets. 

\section{Conclusions}

A model has been presented for APs in a responsive BM with an explicit treatment of the phase separation medium, acting as a soft confining medium. 
APs interact with the background fluid, which in turn, can modify its equilibrium droplet morphology due to activity, forming emulsions. 
The effect of confinement has been quantified in terms of dimensionless parameters resulting from the competition between active energy and surface tension. 

A rich state behaviour has been identified in terms of the AP concentration and the activity rate where different regimes can be differentiated based on the droplet morphology and the AP localisation. 
APs can break the droplet morphology in two different ways: 
for moderate activities groups of APs can deform the BM interface and stabilise small droplets,
while for highly active particles can deform the interface at a single particle level.

Within the droplet regime it is possible to observe large shape fluctuations along the droplet interface, due to the active pressure exerted by the particles near the droplet boundary. 
Future work will study the power spectrum of the shape fluctuations in order to discern the out-of-equilibrium nature of shape fluctuations\cite{kokot_spontaneous_2022}, 
and explore the ability of APs to induce propulsion in the hosting droplets\cite{gandikota_rectification_2023}.
Furthermore, various mechanism of extrusion\cite{paraschiv_influence_2021} and interface instabilities 
\cite{deluca_supramolecular_2024} motivate more detailed studies on the breakage mechanism detailed in Figs.\ref{fig:break.L256}  and \ref{fig:break.L256.Pe30}.




\section*{Funding}

J.D. acknowledges financial support from the Spanish Ministry of Universities through the Recovery, Transformation and Resilience Plan funded by the European Union (Next Generation EU), and Universitat de Barcelona.
I.P. acknowledges support from Ministerio de Ciencia, Innovaci\'on y
Universidades MCIU/AEI/FEDER for financial support under
grant agreement PID2021-126570NB-100 AEI/FEDER-EU, from
Generalitat de Catalunya  under Program Icrea Acad\`emia and project 2021SGR-673.








\bibliographystyle{tfo}

\begin{thebibliography}{32}
\providecommand{\url}[1]{\texttt{#1}}
\providecommand{\urlprefix}{URL }

\bibitem{bechinger_active_2016}
C. Bechinger, R. Di~Leonardo, H. Löwen, C. Reichhardt, G. Volpe and G. Volpe,  Reviews of Modern Physics  \textbf{88} (4), 045006 (2016).

\bibitem{woodhouse_spontaneous_2012}
F.G. Woodhouse and R.E. Goldstein,  Physical Review Letters  \textbf{109} (16), 168105 (2012).

\bibitem{wioland_confinement_2013}
H. Wioland, F.G. Woodhouse, J. Dunkel, J.O. Kessler and R.E. Goldstein,  Physical Review Letters  \textbf{110} (26), 268102 (2013).

\bibitem{lushi_fluid_2014}
E. Lushi, H. Wioland and R.E. Goldstein,  Proceedings of the National Academy of Sciences  \textbf{111} (27), 9733--9738 (2014).

\bibitem{reichhardt_pattern_2023}
C. Reichhardt and C.J.O. Reichhardt,  Europhysics Letters  \textbf{142} (3), 37001 (2023).

\bibitem{wioland_ferromagnetic_2016}
H. Wioland, F.G. Woodhouse, J. Dunkel and R.E. Goldstein,  Nature Physics  \textbf{12} (4), 341--345 (2016).

\bibitem{fernandez-rodriguez_feedback-controlled_2020}
M.A. Fernandez-Rodriguez, F. Grillo, L. Alvarez, M. Rathlef, I. Buttinoni, G. Volpe and L. Isa,  Nature Communications  \textbf{11} (1), 4223 (2020).

\bibitem{bischofberger_editorial_2023}
I. Bischofberger, R. Castañeda-Priego, R. Cerbino, L. Cipelletti, E.D. Gado and A. Fernandez-Nieves,  Frontiers in Physics  \textbf{11} (2023).

\bibitem{needleman_active_2017}
D. Needleman and Z. Dogic,  Nature Reviews Materials  \textbf{2} (9), 1--14 (2017).

\bibitem{aranson_bacterial_2022}
I.S. Aranson,  Reports on Progress in Physics  \textbf{85} (7), 076601 (2022).

\bibitem{fang_nonequilibrium_2019}
X. Fang, K. Kruse, T. Lu and J. Wang,  Reviews of Modern Physics  \textbf{91} (4), 045004 (2019).

\bibitem{takatori_active_2020}
S.C. Takatori and A. Sahu,  Physical Review Letters  \textbf{124} (15), 158102 (2020).

\bibitem{peterson_vesicle_2021}
M.S.E. Peterson, A. Baskaran and M.F. Hagan, Nature Communications \textbf{12} (1) (2021).

\bibitem{paoluzzi_shape_2016}
M. Paoluzzi, R. Di~Leonardo, M.C. Marchetti and L. Angelani,  Scientific Reports  \textbf{6} (1), 34146 (2016).

\bibitem{li_shape_2019}
Y. Li and P.R. ten Wolde,  Physical Review Letters  \textbf{123} (14), 148003 (2019).

\bibitem{vutukuri_active_2020}
H.R. Vutukuri, M. Hoore, C. Abaurrea-Velasco, L. van Buren, A. Dutto, T. Auth, D.A. Fedosov, G. Gompper and J. Vermant,  Nature  \textbf{586} (7827), 52--56 (2020).

\bibitem{iyer_dynamic_2023}
P. Iyer, G. Gompper and  D. A. Fedosov, Soft Matter \textbf{19} (19),3436-3449 (2023).

\bibitem{kokot_spontaneous_2022}
G. Kokot, H.A. Faizi, G.E. Pradillo, A. Snezhko and P.M. Vlahovska,  Communications Physics  \textbf{5} (1), 1--7 (2022).

\bibitem{diaz_emergent_2023}
J. D\'{i}az and I. Pagonabarraga,  Physical Review E  \textbf{108} (6), L062601 (2023).

\bibitem{pinna_modeling_2011}
M. Pinna, I. Pagonabarraga and A.V. Zvelindovsky,  Macromolecular Theory and Simulations  \textbf{20} (8), 769--779 (2011).

\bibitem{diaz_hybrid_2022}
J. Diaz, M. Pinna, A.V. Zvelindovsky and I. Pagonabarraga,  Polymers  \textbf{14} (9), 1910 (2022).

\bibitem{tanaka_simulation_2000}
H. Tanaka and T. Araki,  Physical Review Letters  \textbf{85} (6), 1338--1341 (2000).

\bibitem{cahn_free_1959}
J.W. Cahn,  The Journal of Chemical Physics  \textbf{30} (5), 1121--1124 (1959).

\bibitem{cahn_free_1959-1}
J.W. Cahn and J.E. Hilliard,  The Journal of Chemical Physics  \textbf{31} (3), 688--699 (1959).

\bibitem{cook_brownian_1970}
H.E. Cook,  Acta Metallurgica  \textbf{18} (3), 297--306 (1970).

\bibitem{ball_spinodal_1990}
R.C. Ball and R.L.H. Essery,  Journal of Physics: Condensed Matter  \textbf{2} (51), 10303--10320 (1990).

\bibitem{oono_computationally_1987}
Y. Oono and S. Puri,  Physical Review Letters  \textbf{58} (8), 836--839 (1987).

\bibitem{ohta_anomalous_1993}
T. Ohta, Y. Enomoto, J.L. Harden and M. Doi,  Macromolecules  \textbf{26} (18), 4928--4934 (1993).

\bibitem{solon_pressure_2015}
A.P. Solon, J. Stenhammar, R. Wittkowski, M. Kardar, Y. Kafri, M.E. Cates and J. Tailleur,  Physical Review Letters  \textbf{114} (19), 198301 (2015).

\bibitem{speck_ideal_2016}
T. Speck and R.L. Jack,  Physical Review E  \textbf{93} (6), 062605 (2016).


\bibitem{howse_self-motile_2007}
J.R. Howse, R.A.L. Jones, A.J. Ryan, T. Gough, R. Vafabakhsh and R. Golestanian,  Physical Review Letters  \textbf{99} (4), 048102 (2007).

\bibitem{li_accumulation_2009}
G. Li and J.X. Tang,  Physical Review Letters  \textbf{103} (7), 078101 (2009).

\bibitem{palacios_guided_2021}
L.S. Palacios, S. Tchoumakov, M. Guix, I. Pagonabarraga, S. Sánchez and A. G.~Grushin,  Nature Communications  \textbf{12} (1), 4691 (2021).

\bibitem{gandikota_rectification_2023}
M.S. Gandikota and A. Cacciuto, Soft Matter \textbf{19} (2), 315-320 (2023)

\bibitem{paraschiv_influence_2021}
A. Paraschiv, T. J. Lagny, E. Coudrier, C. V. Campos, P. Bassereau and A. Šarić, Biophysical Journal  \textbf{120} (598) (2021).

\bibitem{deluca_supramolecular_2024}
F. De~Luca, I. Maryshev and E. Frey,  arXiv preprint arXiv:2401.05070   (2024).

\bibitem{menzl_molecular_2016}
G. Menzl, M.A. Gonzalez, P. Geiger, F. Caupin, J.L. Abascal, C. Valeriani and C. Dellago,  Proceedings of the National Academy of Sciences  \textbf{113} (48), 13582--13587 (2016).

\end{thebibliography}


\appendix

\section{Additional information}

\begin{figure}
    \centering
    \includegraphics[width=1\linewidth]{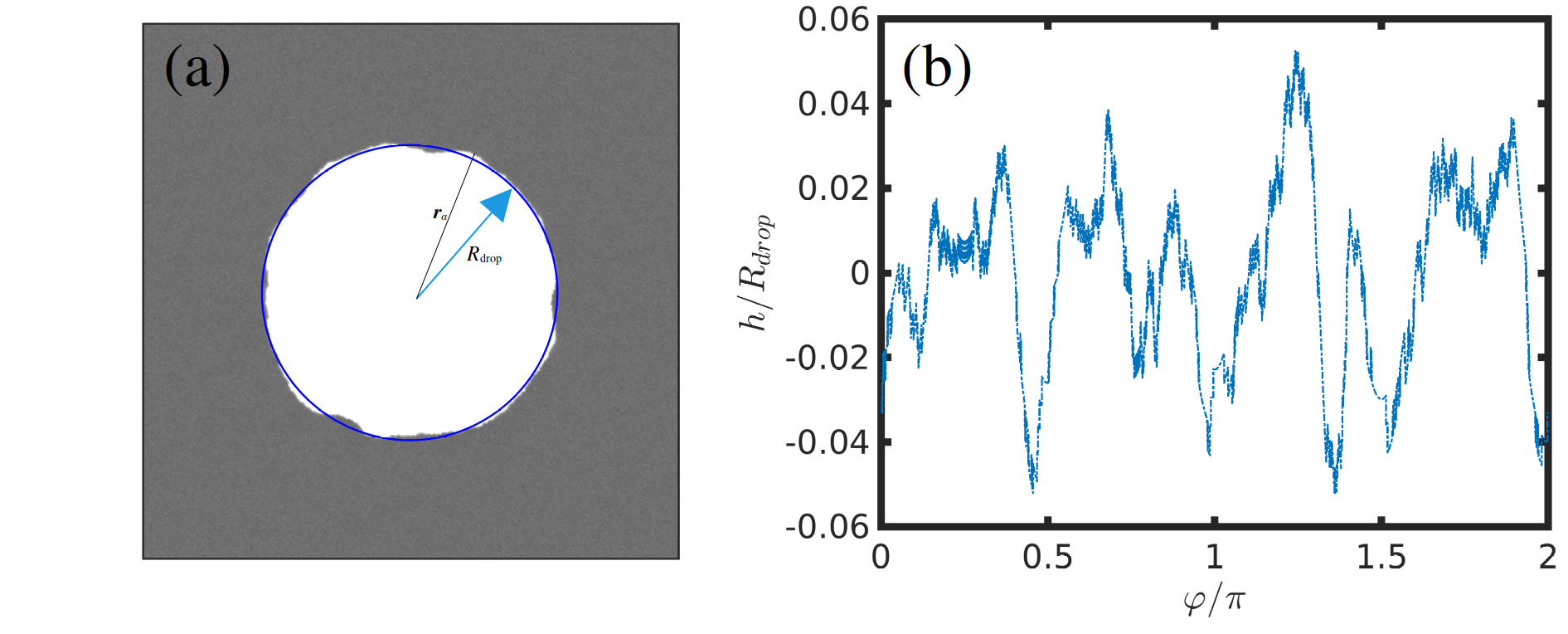}
    \caption{
        Scheme on the determination of the interface points $\rvec_{\alpha}$ and the droplet size $R_{drop}$ in (a) to obtain the droplet fluctuations over the contour of the interface $h(\phi)$. 
    }
    \label{fig:scheme}
\end{figure}

The scheme shown in Fig.  \ref{fig:scheme} illustrates the method to calculate interface fluctuations for a given droplet. 

Table \ref{tab:movies} summarises the movies provided as supporting material to facilitate the visualisation of the dynamic behaviour of the system. 

\begin{table}
    \centering
    \begin{tabular}{ccccc}
         Name & Corr. Fig. & $\phi_p$  &  $Pe$ & Comment \\
         \hline
         S1 & \ref{fig:msd} (c) & $0.01$  & $10^{-1}$  & Dilute regime under confinement \\
         S2 & \ref{fig:msd} (c) & $0.01$  & $10^0$  & Dilute regime under confinement \\
         S3 & \ref{fig:msd} (c) & $0.01$  & $10^{1}$  & Dilute regime under confinement \\
         S4 & \ref{fig:msd} (c) & $0.01$  & $10^{2}$  & Dilute regime under confinement \\
         S5 & \ref{fig:break.L256} & $0.1$  & $10$  & Droplet breakage \\
         S6 & \ref{fig:break.L256.Pe30} & $0.1$  & $30$  & Droplet breakage \\
    \end{tabular}
    \caption{
        Summary of movies associated with corresponding figures in the text. 
    }
    \label{tab:movies}
\end{table}

\begin{figure}
    \centering
    \includegraphics[width=0.5\linewidth]{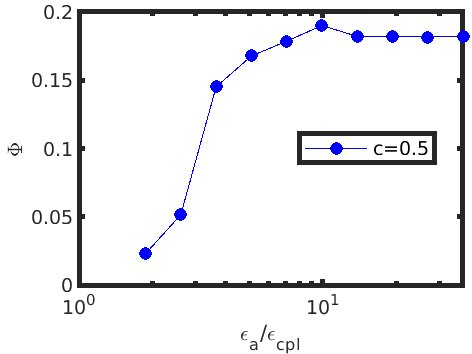}
    \caption{
        Fraction of APs in the white phase in the limit of large $Pe$ for $c=0.5$. 
    }
    \label{fig:fraction.Peinf}
\end{figure}

Fig. \ref{fig:fraction.Peinf} shows the fraction of APs within the white phase when $Pe$ is explored up to large values while $c=0.5$. 

\section{Asphericity of droplet due to activity}

In order to quantify the departure from isotropic shape of droplets containing APs, we calculate the asphericity of the droplet with the expression\cite{menzl_molecular_2016}
\begin{equation}
    \alpha=\lambda_1/\lambda_2 -1
\end{equation}
where $\lambda_1$ and $\lambda_2<\lambda_1  $ are the two eigenvalues of the radius of gyration tensor $S_{\alpha \beta}$ obtained from the domain analysis detailed in the main text. 
The $\alpha,\beta$ components of the gyration tensor can be calculated as 
\begin{equation}
    S_{\alpha \beta} = 
    \frac{1}{N}\sum_i \Delta\rvec_i^{\alpha} \Delta\rvec_{i}^{\beta} 
\end{equation}
where $\Delta\rvec_i^{\alpha}$ is the $\alpha=x,y$ component of the interface point $i=1...N$ which specify the border of a droplet, with respect to the centre of mass of the droplet.  
We note that when two or more droplets are present in the system (see Fig. \ref{fig:curves.L256} (b) in the main text for $N>1$) the asphericity of the droplets is determined as the average over all droplets.

\begin{figure}
    \centering
    \includegraphics[width=0.5\linewidth]{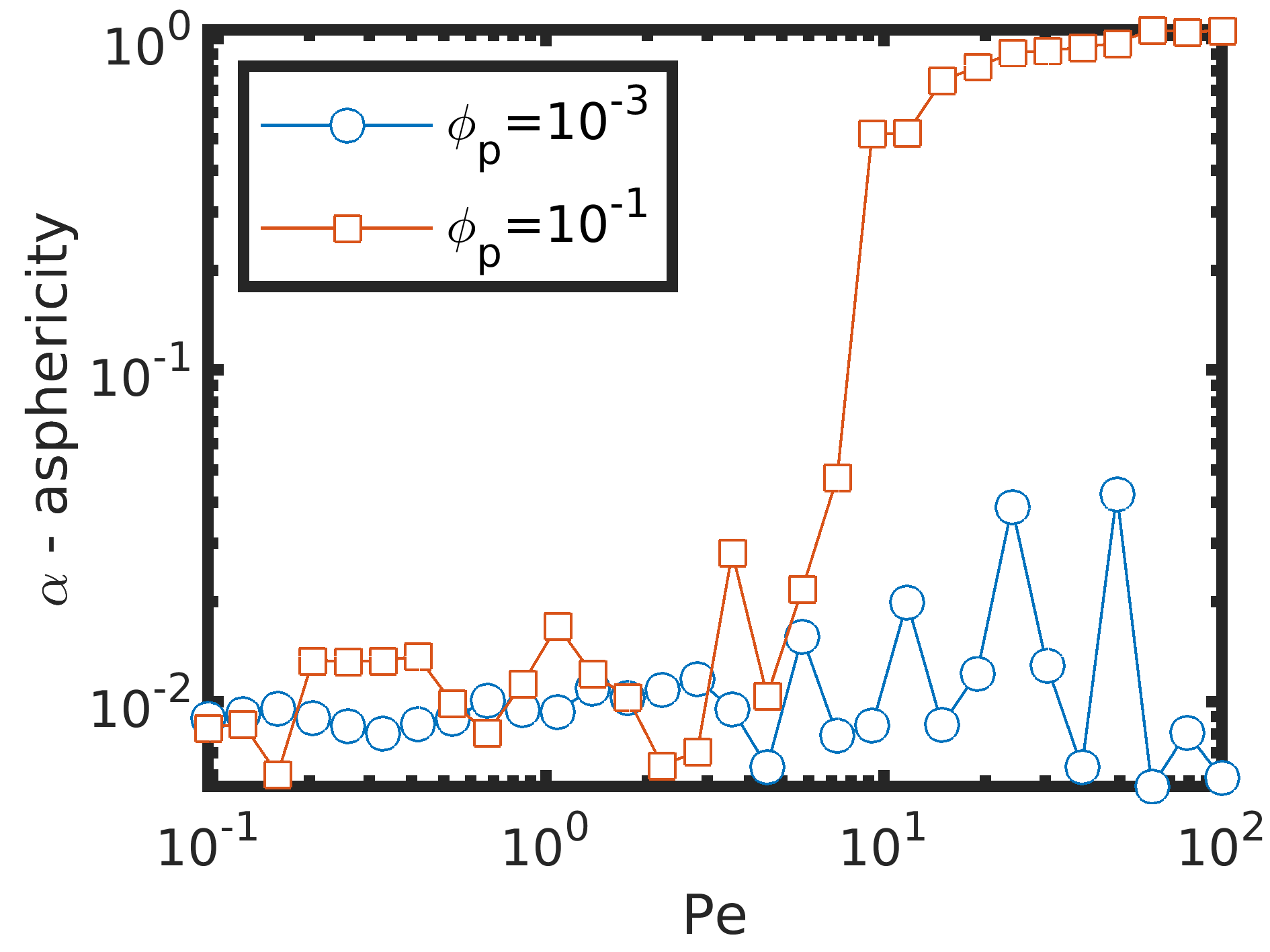}
    \caption{
        Asphericity of a droplet with size $R_{drop}=18.9$ containing APs with activity given by $Pe$.  
    }
    \label{fig:asphere}
\end{figure}

\section{Colourmaps used for state diagrams}

\begin{figure}
    \centering
    \includegraphics[width=1\linewidth]{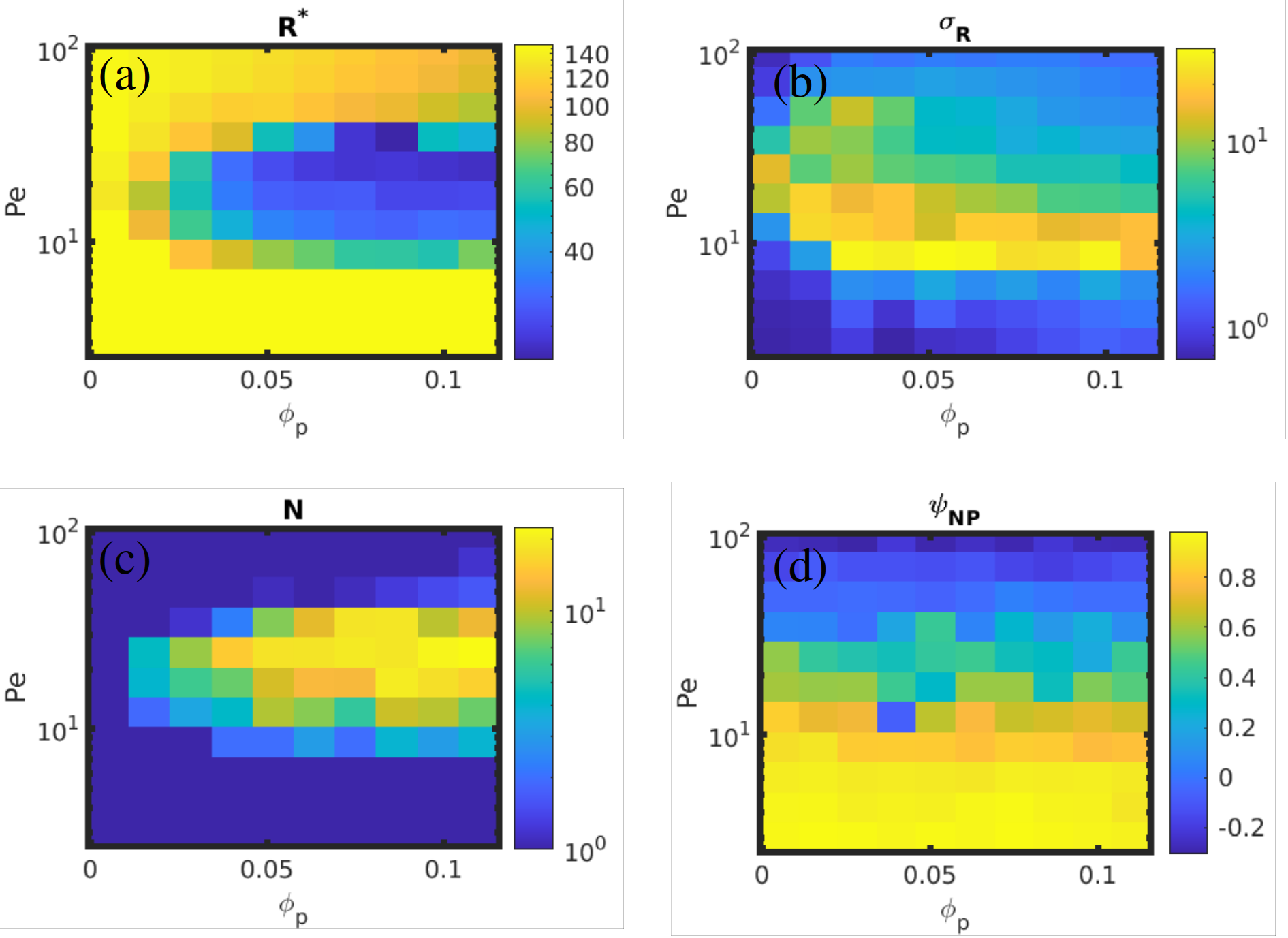}
    \caption{
        Colourmaps of observables used for identifying the regimes in state diagram in Fig. \ref{fig:phd.L256} for a droplet with size $R_{drop}=64$. 
        (a) shows the steady-state characteristic size of the BM $R^*$, 
        (b) shows the standard deviation of the droplet size $\sigma_R$, 
        (c) shows the number of BM domains $N$, and 
        (d) shows the average value of $\psi$ in the vicinity of the APs $\psi_{NP}$.  
    }
    \label{fig:phd.L256.colormaps}
\end{figure}

Fig.~\ref{fig:phd.L256.colormaps} shows the colourmaps of the observables used to characterise the state diagram in Fig.~\ref{fig:phd.L256}. 

\section{Effect of BM composition}

\begin{figure}
    \centering
    \includegraphics[width=0.5\linewidth]{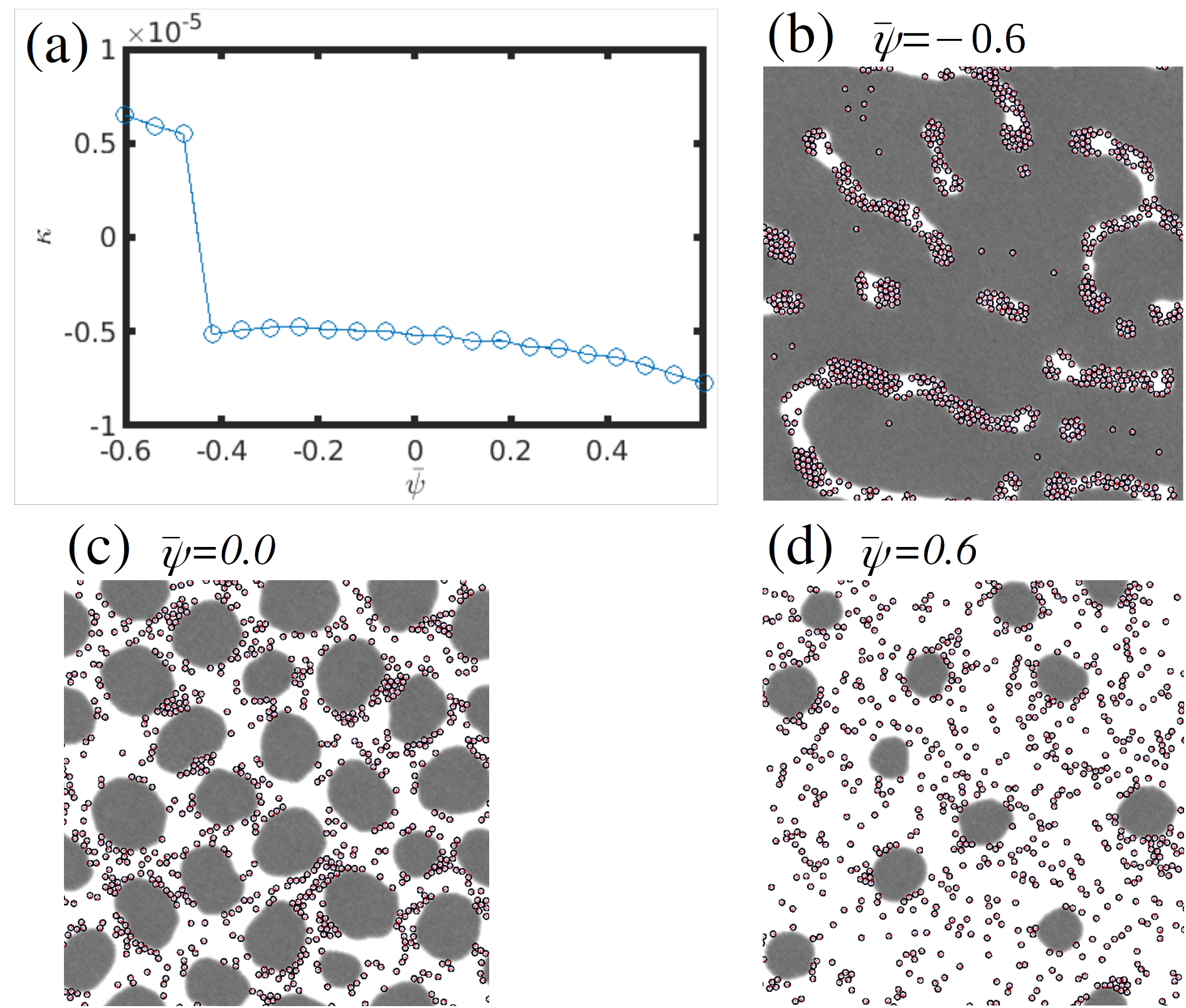}
    \caption{
        Role of BM composition $\barpsi$ for fixed $Pe=10$ and $\phi_p=0.1$. 
        The BM and the APs are initialised as a disordered state. 
    }
    \label{fig:explore.psimean}
\end{figure}

In this work we have focused on a fixed BM composition $\barpsi = -0.61$ in order to generate well-defined droplets enclosing APs.
However, we can explore the role of the BM composition by modifying $\barpsi$ as in Fig.~\ref{fig:explore.psimean} in regime II with $\phi_p=0.1$ and $Pe=10$. 

In Fig. \ref{fig:explore.psimean} we find that activity stabilises non-equilibrium morphologies different from macroscopic phase separation. 
In order to characterise the morphology of the BM we calculate the mean curvature of the  BM interface by defining a local normal unit vector to the interface $\hat{\textbf{n}} = \frac{\nabla \psi}{||\nabla \psi||}$ where $|| * ||$ is the modulus of a vector. 
The tangential vector is $\hat{\textbf{t}}$ and its change can be calculated with the tensor 
$
\bar{\bar{ C }} = \nabla  \hat{\textbf{t}} 
$
with components $ C_{ij} = \frac{\partial t_i}{\partial x_j}$ . 
Its projection into the tangential direction is related to the local curvature
$\textbf{k}(\textbf{r}) = \bar{\bar{ C }}\cdot \hat{\textbf{t}}$. 
Finally, the total curvature is averaged over space with a weight $||\nabla \psi ||$ to minimise the bulk contributions. 

\section{Passive droplet shape fluctuations}

As a reference point, we consider the thermal shape fluctuations in the absence of activity. 
This can be understood as the limit in which $Pe\to 0$ in the main work. 

\begin{figure}
    \centering
    \includegraphics[width=0.5\linewidth]{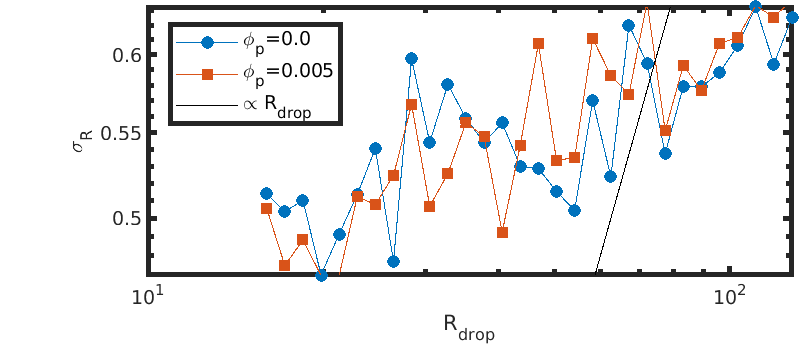}
    \caption{Passive fluctuations in the $Pe=0$ limit for droplet size $R_{drop}$ with and without particles.}
    \label{fig:passive-fluc}
\end{figure}

Fig. \ref{fig:passive-fluc} shows the fluctuations of the droplet shape depending on the droplet size $R_{drop}$ for two regimes: 
no particles $\phi_p=0.0$ and small concentration of particles $\phi_p=0.005$.
In both cases the fluctuations are shown to scale weakly with the droplet size. 
For reference, a linear scaling is shown as a solid line. 

\end{document}